\documentclass[english]{llncs}

\usepackage[top=5.2cm, bottom=5.2cm, left=4.4cm, right=4.4cm]{geometry}

\usepackage[T1]{fontenc}
\usepackage{times} 
\usepackage{graphicx} 
\usepackage{tabularx}
\usepackage{authblk}
\usepackage{csquotes}
\usepackage{orcidlink}
\usepackage{xcolor}
\usepackage{threeparttable}
\usepackage{url}
\usepackage[english,spanish]{babel}
\usepackage[style=apa, sorting=ydnt]{biblatex}

\date{} 

\begin{document}

\title{Qiskit Code Migration with LLMs}

\author{José Manuel Suárez\inst{1,2}\orcidlink{0009-0001-1115-6225} \and
Luís Mariano Bibbó\inst{1,2}\orcidlink{0000-0003-4950-3605} \and
Joaquín Bogado\inst{1,2}\orcidlink{0000-0001-9491-5698}  \and
Alejandro Fernandez\inst{1,2,3}\orcidlink{0000-0002-7968-6871}}

\authorrunning{F. Author et al.}

\institute{Laboratorio de Investigación y Formación en Informática Avanzada (LIFIA) \and
Facultad de Informática - Universidad Nacional de La Plata, Buenos Aires, Argentina  \and
Comisión de Investiación Científica - CIC Bs.As.}

\maketitle

\selectlanguage{english}
\begin{abstract}

The rapid evolution of Quantum Development Kits (QDKs) introduces a specific form of technical debt that compromises code maintainability and hinders software reuse. In the specialized domain of Quantum Software Engineering (QSE), this challenge is intensified by the scarcity of high-quality training data and the high volatility of emerging frameworks, which often lead general-purpose Large Language Models (LLMs) to produce unreliable or hallucinated results. This paper proposes a hybrid approach integrating LLMs with Retrieval-Augmented Generation (RAG) to automate the migration of Qiskit code across versions. The proposed methodology enhances the precision and reliability of migration suggestions by leveraging an automatically generated taxonomy of migration scenarios as the structured, version-specific knowledge source to guide the models. The approach is implemented through an automated, extensible workflow evaluating LLMs (Google Gemini Flash-2.5 and OpenAI Gpt-oss-20b) under different retrieval schemes (unconstrained and restrictive). Results demonstrate that the taxonomy-based RAG architecture, particularly under the restrictive scheme, significantly reduces hallucinations and improves descriptive quality, with Google Gemini Flash-2.5 showing superior performance in detecting complex refactoring scenarios. These findings confirm the potential of this data-centric methodology to foster technological independence and provide robust, intelligent assistants that mitigate API obsolescence, ensuring the long-term availability of quantum algorithms within a rapidly shifting ecosystem and flattening the learning curve within Quantum Software Engineering (QSE).

\vspace{1em}

\end{abstract}

\vspace{0.5em}

\section{Introduction}
Driven by hardware innovations, simulation, and optimized techniques in error correction and fault tolerance (\cite{wildon_quantum_2026, adermann_quantum_2026, aasen_roadmap_2025}), the rapid progress of QC is accelerating the evolution of QDKs (\cite{ahmad_quantum_2025, gill_quantum_2025}). This expansion has diversified the QC community beyond physics and mathematics to include fields like computer science, biology, and engineering. However, this interdisciplinary shift imposes significant challenges on development tools, particularly regarding accessibility and the learning curve. Moreover, it highlights the need to address legacy issues from classical software engineering (CSE), such as API obsolescence triggered by disruptive version releases.

The convergence of QC and CSE has given rise to an emerging discipline: quantum software engineering (QSE) (\cite{khan_advancing_2025, mandal_quantum_2025, murillo_quantum_2025}), a field that adapts empirical methodologies to the software lifecycle of hybrid quantum-classical systems within NISQ-era constraints (\cite{lammers_quantum_2025, lau_nisq_2022, preskill_quantum_2018}). Concurrently, AI agents—specifically Large Language Models (LLMs) (\cite{minaee_large_2025, xiao_foundations_2025})—have proven effective in managing complex tasks, prompting a fundamental rethink of workflows and tools within the discipline.

Previous stages demonstrated the constructive feasibility of a taxonomy that consolidates migration scenarios (\cite{suarez_taxonomy_2025}) and established its efficacy in guiding LLMs for scenario detection and the generation of refactoring suggestions (\cite{suarez_automatic_2025}). The present work integrates: (i) generative artificial intelligence (GenAI) elements (\cite{sengar_generative_2024, liang_foundations_2024, weisz_design_2024}), through the configurable download and execution of LLMs; (ii) low-code/no-code (LCNC) tools \footnote{IBM LCNC \url{https://www.ibm.com/think/topics/low-code}}, aiming to automate the experimental workflow and ensure its extensibility; (iii) a retrieval-augmented generation (RAG) architecture (\cite{gupta_comprehensive_2024, gao_retrieval-augmented_2024, mombaerts_meta_2024}), to configure a knowledge base that complements the models' generic information; and (iv) the traceability of our experimental base, supported by the semi-automatic execution of scenario detection and suggestion generation tasks, as well as code migration and metrics analytics previously developed for the Qiskit QDK (\cite{github_qiskitSDK_2025, pathak_evolution_2025, aragones-soria_architecture_2025}) \footnote{IBM Qiskit \url{https://www.ibm.com/quantum/qiskit}}.

API obsolescence and the resulting technical gap are critical challenges in CSE, particularly in QC due to the accelerated evolution of QDKs, which threatens algorithmic functionality and service availability (\cite{panter_technical_2026, wang_how_2024, wang_searching_2024, chen_every_2026, liang_foundations_2024}). While LLMs streamline development, they face reliability issues in data-scarce emerging domains like QC (\cite{dobariya_mind_2025, zheng_enhancing_2025, chen_quantifying_2023}). This research develops AI assistants to address QSE challenges related to API obsolescence and quantum algorithm availability, aiming for agile adaptation and robust migration criteria in volatile ecosystems (\cite{aragones-soria_c4q_2024}).

\begin{figure}[htbp]
    \centerline{\includegraphics[width=1\columnwidth]{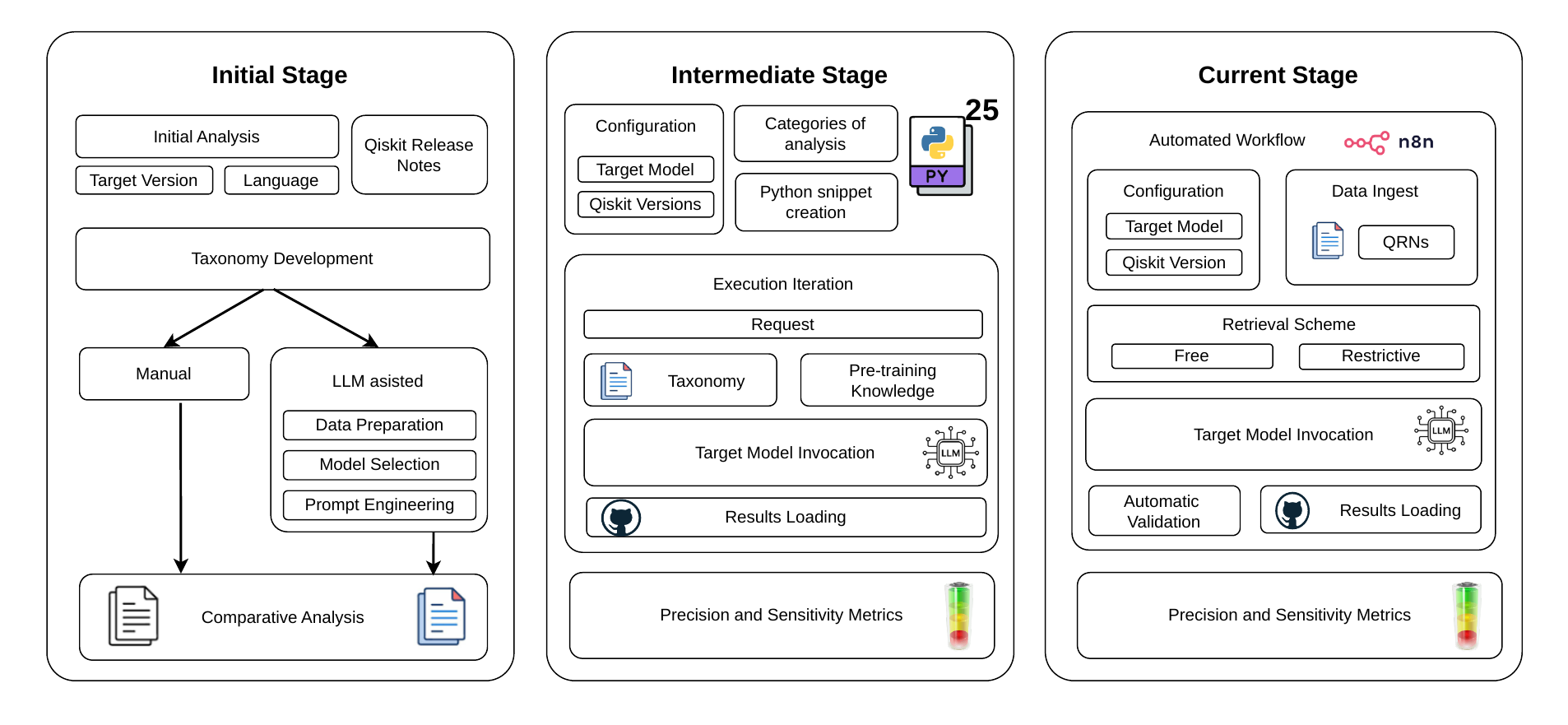}}
    \caption{Stages of our research line}
    \label{fig:etapas_linea_investigativa}
\end{figure}

Once the components of the solution-oriented approach were defined, the continuity of our research line was outlined, as illustrated in Figure~\ref{fig:etapas_linea_investigativa}. In the initial stage (\cite{suarez_taxonomy_2025}), we evaluated the utility of a taxonomy designed to consolidate authoritative information on API evolution following disruptive (major) Qiskit releases. This stage employed a hybrid constructive approach (manual and LLM-automated) to establish comparative baselines and verify the potential of LLMs. In the subsequent stage (\cite{suarez_automatic_2025}), metrics were generated to compare effectiveness in detecting migration scenarios and the precision of adaptation suggestions, using a battery of synthetic Python code snippets \footnote{Python Software Foundation - \url{https://www.python.org/}}. In the current stage, the effort focuses on integrating the automated migration scenario taxonomy into a RAG architecture to complement the models' general knowledge with contextually relevant, authoritative, and updated information. Additionally, this approach is channeled through an automated workflow that streamlines the experimental framework.

This work contributes: (i) a taxonomy-based RAG architecture using semantic databases; (ii) an automated, collaborative workflow; (iii) an impact analysis of different retrieval schemes; and (iv) automated code validation to augment expert review. The rest of this paper is structured as follows: Section 2 reviews related work; Section 3 details the methodology; Section 4 presents the results; Section 5 discusses the findings; and Section 6 concludes with future research directions.

\section{Related Work}

Regarding the intersection of code refactoring, RAG and LLMs, \cite{bellur_together_2025} introduced MM-ASIST, an IDE-integrated assistant. However, their study focuses exclusively on 'Move Method' refactoring. \cite{hostnik_retrieval-augmented_2025} compared RAG strategies for Python code completion using small local models to prioritize privacy. While this work also addresses reliability, custom metrics and expert evaluation are utilized, noting that small models may prove insufficient for QSE complexities. \cite{closser_pushing_2025} evaluated four public LLMs for quantum circuit synthesis, identifying Gemini—utilized in these experiments—as the most efficient. \cite{siavash_model-driven_2025} explored Qiskit code generation from UML via MDE and RAG. Although they aim to mitigate hallucinations through authoritative repositories, their work does not address refactoring. Finally, \cite{li_mitigating_2025} investigated strategies to enhance LLM capabilities and reduce hallucinations. This approach can be seen as a domain-specific instance of their proposed schemes, though further study is required to fully evaluate the impact of specialized prompting within QSE.

\cite{henderson_programming_2025} evaluated LLMs' ability to write Qiskit code to flatten the learning curve in QC. While general models may suffice for basic algorithm generation, we argue that migrating existing code across specific versions requires a RAG-based architecture with domain-specific knowledge. \cite{campbell_enhancing_2025} compared Chain-of-Thought (CoT) strategies with RAG for Qiskit-based algorithm generation and error correction. They noted that standard RAG often fails due to outdated information—a limitation our version-specific taxonomy architecture specifically addresses.
Other approaches include fine-tuning for high-quality Qiskit code generation \cite{dupuis_qiskit_2024} and specialized data generation for Pennylane \cite{asif_pennylang_2025}. In the industry, Microsoft's Azure Quantum Copilot incorporates RAG but does not target refactoring or version migration. Similarly, \cite{kheiri_qspark_2025} introduced QSpark, a fine-tuned assistant for generating Qiskit code from scratch. Unlike these model-centric approaches, our work is data-centric, focusing on taxonomies to refactor existing code, thereby avoiding the complexity and cost of fine-tuning. Finally, \cite{abufarha_mitigating_2026} proposed a hybrid assistant to reduce prompting dependency in refactoring and testing. While these studies share contact points with our research, they do not specifically address code migration within the context of API obsolescence in QSE.

\section{Methodology}

The experimental design centers on constructing a semantic database through a retrieval technique that provides models with automatically generated, structured information (\cite{minaee_large_2025, yuan_transagent_2024}). This approach offers a pragmatic alternative to relying solely on the model's internal knowledge and avoids costly, complex strategies such as fine-tuning (\cite{gao_retrieval-augmented_2024}). By incorporating RAG, we aim to steer retrieval by prioritizing reliable data from the underlying taxonomies, thereby reducing hallucinations and improving refactored code accuracy. To operationalize this, we implemented a highly flexible workflow using n8n \footnote{n8n tool - \url{https://docs.n8n.io/}}, based on a GitHub-hosted project \footnote{GitHub - \url{https://docs.github.com/en}} developed in Visual Studio Code \footnote{Visual Studio Code IDE - \url{https://code.visualstudio.com/docs}}. This workflow encapsulates the experimental framework, enabling the execution of both local and remote models, pre- and post-processing stages, integration with external services, and the automatic uploading of results to the source repository.

\begin{figure}[htbp]
    \centerline{\includegraphics[width=1\columnwidth]{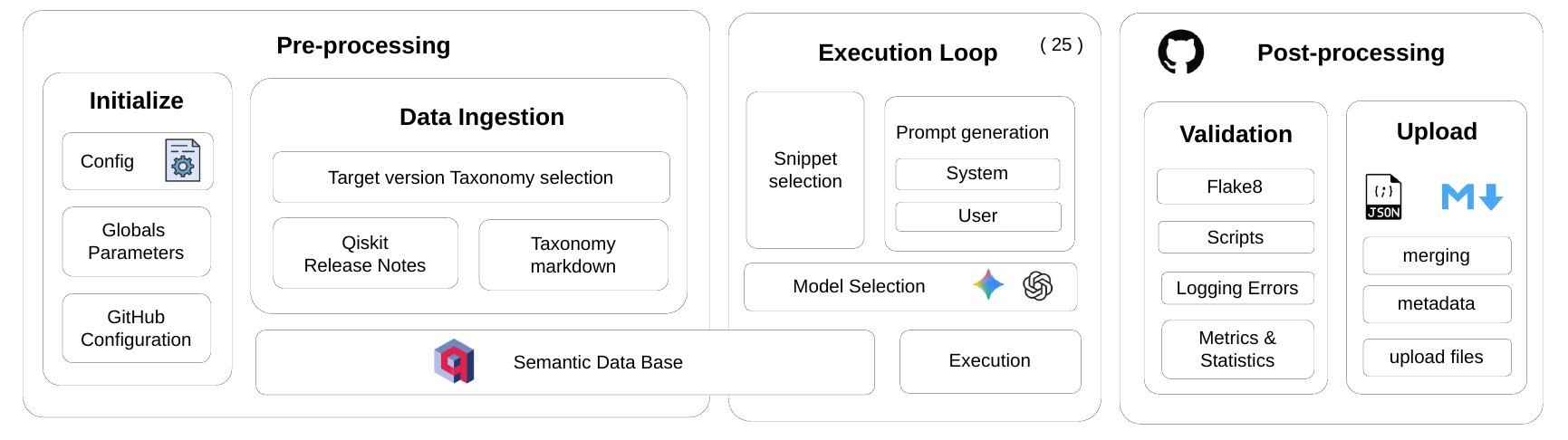}}
    \caption{Workflow stages in n8n}
    \label{fig:n8n_workflow}
\end{figure}

Using n8n aims to enhance agility, flexibility, and scalability while reducing technical overhead (\cite{pattnayak_review_2025, heuschkel_impact_2023}). The experimental workflow includes the following key stages (see Figure~\ref{fig:n8n_workflow}): (i) Global Configuration, which defines functional parameters to adapt experimental schemes, including the semantic database, test models, GitHub repository, Qiskit version, retrieval strategy, and verification stages; (ii) Data Ingestion, which extracts release notes and associated automated taxonomies from the source repository into a semantic database; (iii) Data Processing, which handles test snippets and prompts—generated in either free or restrictive mode—and establishes tracking metadata; (iv) Request Execution, the core loop where each iteration loads a test scenario, prompts, and queries the target model, while the RAG mechanism manages access to both local data and pre-acquired knowledge; (v) Automated Validation; and (vi) Results Storage, which automatically uploads experimental results.

The models used were Open AI Gpt-oss-20b (\cite{noauthor_presentamos_2026, openai_gpt-4_2024}) and Google Gemini Flash-2.5 (\cite{comanici_gemini_2025, team_gemini_2024}), keeping default configurations. The official information sources were Qiskit Release Notes \footnote{Qiskit Release Notes - \url{https://quantum.cloud.ibm.com/docs/en/api/qiskit/release-notes}}, as the basis for generating the automatic taxonomy of migration scenarios. Regarding the prompts, a one-shot strategy was used. 
In the post-analysis stage, the detected scenarios were quantified and the accuracy of recommendations was classified using our metric, allowing for comparative contrast in the line of research.
We decided to work with two retrieval schemes: (i) a free scheme, where the requests made to the model are limited to experimental functional requirements without biasing the model on the retrieval of information; and (ii) a restrictive scheme, where, in addition to functional requirements, the model is directed towards the retrieval of information from the Qdrant semantic database \footnote{Qdrant Database - \url{https://qdrant.tech/documentation/}}.

\section{Results}

Following the proposed experimental pipeline and methodology, results are summarized in Tables \ref{tabla:1} and \ref{tabla:2}. Evaluation employed the stoplight metrics established by \cite{suarez_automatic_2025}. Two experts and a Qiskit engineer worked independently using a double-blind technique. The test models, Google Gemini Flash-2.5 and OpenAI Gpt-oss-20b, were evaluated according to the workflow described previously. 
The distribution of scenarios returned by each model was analyzed based on the imposed retrieval schemes. Detailed scenario distributions according to the stoplight metrics are presented in Table \ref{tabla:2}.

\begin{table}[htbp]
    \centering
    \scriptsize 
    \renewcommand{\arraystretch}{1.5} 
    \setlength{\tabcolsep}{3pt} 
    
    \caption{Summary of results for test models and information retrieval modes}
    \label{tabla:1}
    
    \begin{tabularx}{\textwidth}{|c|>{\raggedright\arraybackslash}X|*{4}{>{\centering\arraybackslash}p{1.2cm}|}}
    \hline
    \textbf{Case} & \textbf{Description} & \multicolumn{2}{c|}{\textbf{Gemini Flash-2.5}} & \multicolumn{2}{c|}{\textbf{Gpt-oss-20b}} \\
    \cline{3-6}
    & & \textbf{Free} & \textbf{Restrictive} & \textbf{Free} & \textbf{Restrictive} \\
    \hline
    X & Incorrect scenario and inadequate suggestion & 16 & 15 & 13 & 13 \\
    X+ & Incorrect scenario or inadequate suggestion & 4 & 5 & 4 & 3 \\
    OK- & Suggestion needs adaptation & 5 & 9 & 11 & 14 \\
    OK & Correct scenario and suggestion & 80 & 81 & 31 & 37 \\
    P perf. & Expected scenarios detected & 16 & 16 & 9 & 10 \\
    P glob. & Detected vs. expected scenarios & 85 [89] & 90 [89] & 42 [89] & 51 [89] \\[-4pt]
    & (ratio) & (0.95) & (1.01) & (0.47) & (0.57) \\
    \hline
    \multicolumn{2}{|l|}{\textbf{Total detected scenarios}} & 105 & 110 & 59 & 67 \\
    \hline
    \end{tabularx}
\end{table}

\begin{table}[htbp]
    \centering
    \small
    \renewcommand{\arraystretch}{1.3} 
    \setlength{\tabcolsep}{5pt}
    \caption{Distribution of scenarios according to stoplight metrics}
    \label{tabla:2}
    \begin{tabular}{|l|c|c|c|c|}
    \hline
     & \multicolumn{2}{c|}{\textbf{Free}} & \multicolumn{2}{c|}{\textbf{Restrictive}} \\
    \cline{2-5}
    \textbf{Source} & \textbf{Gemini Flash-2.5} & \textbf{Gpt-oss-20b} & \textbf{Gemini Flash-2.5} & \textbf{Gpt-oss-20b} \\
    \hline
    Taxonomy (Tax) & 83 & 15 & 100 & 50 \\
     & {\scriptsize \color{red!80!black}\textbf{8} \ \color{orange!80!black}\textbf{4} \ \color{yellow!80!black}\textbf{5} \ \color{green!60!black}\textbf{66}} & {\scriptsize \color{red!80!black}\textbf{1} \ \color{orange!80!black}\textbf{2} \ \color{yellow!80!black}\textbf{3} \ \color{green!60!black}\textbf{9}} & {\scriptsize \color{red!80!black}\textbf{13} \ \color{orange!80!black}\textbf{5} \ \color{yellow!80!black}\textbf{8} \ \color{green!60!black}\textbf{74}} & {\scriptsize \color{red!80!black}\textbf{6} \ \color{orange!80!black}\textbf{3} \ \color{yellow!80!black}\textbf{11} \ \color{green!60!black}\textbf{30}} \\
    \hline
    Internal (IK) & 22 & 41 & 10 & 3 \\
     & {\scriptsize \color{red!80!black}\textbf{8} \ \color{orange!80!black}\textbf{0} \ \color{yellow!80!black}\textbf{0} \ \color{green!60!black}\textbf{14}} & {\scriptsize \color{red!80!black}\textbf{12} \ \color{orange!80!black}\textbf{2} \ \color{yellow!80!black}\textbf{8} \ \color{green!60!black}\textbf{19}} & {\scriptsize \color{red!80!black}\textbf{2} \ \color{orange!80!black}\textbf{0} \ \color{yellow!80!black}\textbf{1} \ \color{green!60!black}\textbf{7}} & {\scriptsize \color{red!80!black}\textbf{1} \ \color{orange!80!black}\textbf{0} \ \color{yellow!80!black}\textbf{1} \ \color{green!60!black}\textbf{1}} \\
    \hline
    Balance Tax/IK & 83/22 (79/21\%) & 15/41 (27/73\%) & 100/10 (91/9\%) & 50/3 (94/6\%) \\
    \hline
    \textbf{Total} & 105 & 56 & 110 & 53 \\
     & {\scriptsize \color{red!80!black}\textbf{16} \ \color{orange!80!black}\textbf{4} \ \color{yellow!80!black}\textbf{5} \ \color{green!60!black}\textbf{80}} & {\scriptsize \color{red!80!black}\textbf{13} \ \color{orange!80!black}\textbf{4} \ \color{yellow!80!black}\textbf{11} \ \color{green!60!black}\textbf{28}} & {\scriptsize \color{red!80!black}\textbf{15} \ \color{orange!80!black}\textbf{5} \ \color{yellow!80!black}\textbf{7} \ \color{green!60!black}\textbf{83}} & {\scriptsize \color{red!80!black}\textbf{7} \ \color{orange!80!black}\textbf{3} \ \color{yellow!80!black}\textbf{12} \ \color{green!60!black}\textbf{31}} \\
    \hline
    \end{tabular}
    
    \vspace{8pt}
    \footnotesize
    {\color{red!80!black}$\bullet$} \textbf{X: Incorrect} \quad
    {\color{orange!80!black}$\bullet$} \textbf{X+: Inadequate} \quad
    {\color{yellow!80!black}$\bullet$} \textbf{OK-: Req. adjustments} \quad
    {\color{green!60!black}$\bullet$} \textbf{OK: Correct}
\end{table}

\section{Discussion}
\begin{figure}[htbp]
    \centerline{\includegraphics[width=1\columnwidth]{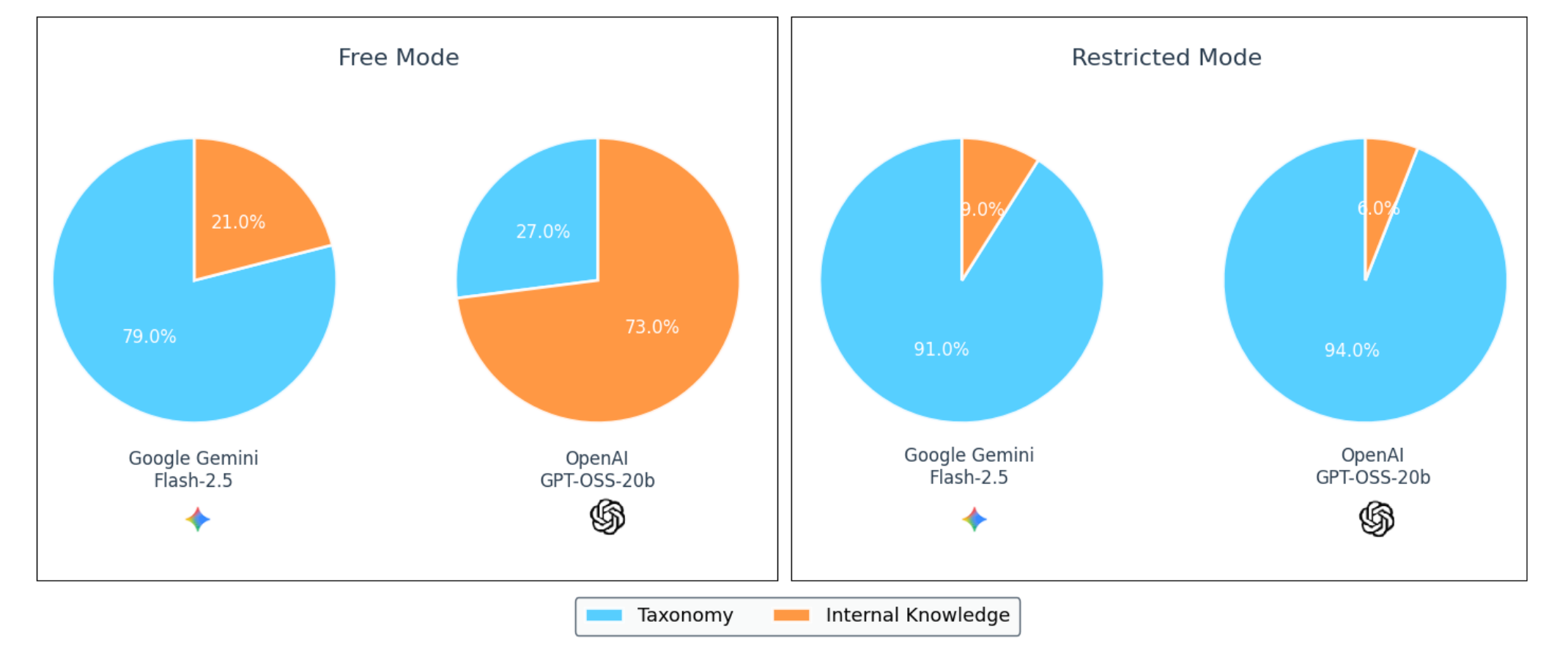}}
    \caption{Retrieval strategies balance}
    \label{fig:strategy_balance}
\end{figure}

Regarding the balance of scenario sources and their distribution in relation to the configured retrieval scheme, \ref{tabla:2} allows us to verify the impact of these strategies on the proposed architecture. For Google Gemini Flash-2.5, a 79/21 ratio is shown in free mode, while in restricted mode the imbalance deepened to 91/9. For OpenAI Gpt-oss-20b, it shifted from 27/73 in free mode to 94/6, also implying a reversal in trend; see Figure~\ref{fig:strategy_balance}

Observations show that Google Gemini Flash-2.5 relies more on authoritative information than OpenAI Gpt-oss-20b, a difference that impacts scenario detection accuracy. Since this margin remains relatively constant regardless of the retrieval strategy, the variation appears to be driven by inherent model characteristics, including parameterization and context window size, rather than the retrieval scheme itself.

The core idea is that the proportion of incorrect refactorings by OpenAI Gpt-oss-20b was moderately reduced from 17 to 16 (6.25\%) when the retrieval balance was reversed, while the proportion of correct scenarios increased from 42 to 51 (21.43\%). Similarly, we interpret that Google Gemini Flash-2.5 has higher performance in terms of detected scenarios (107 on average vs. 81 in previous stages, a 32.1\% increase on average, reaching 35.8\% in our best case under restrictive mode) may be due to the fact that this model, in free mode, resorts more frequently to the taxonomy, allowing it to find more cases, detail them more thoroughly, and provide greater precision in its suggestions.

Based on the results summarized in Table 1, we have verified that Google Gemini Flash-2.5 showed higher performance both in detecting refactoring scenarios and in the quality, precision, and descriptive detail of suggestions, compared to OpenAI gpt-oss-20b, particularly in the number of appropriate scenarios that were directly applicable. On the other hand, no significant changes were reported between the data retrieval strategies, although a slight tendency toward superiority of the restrictive mode is evident.

\begin{figure}[htbp]
    \centerline{\includegraphics[width=1\columnwidth]{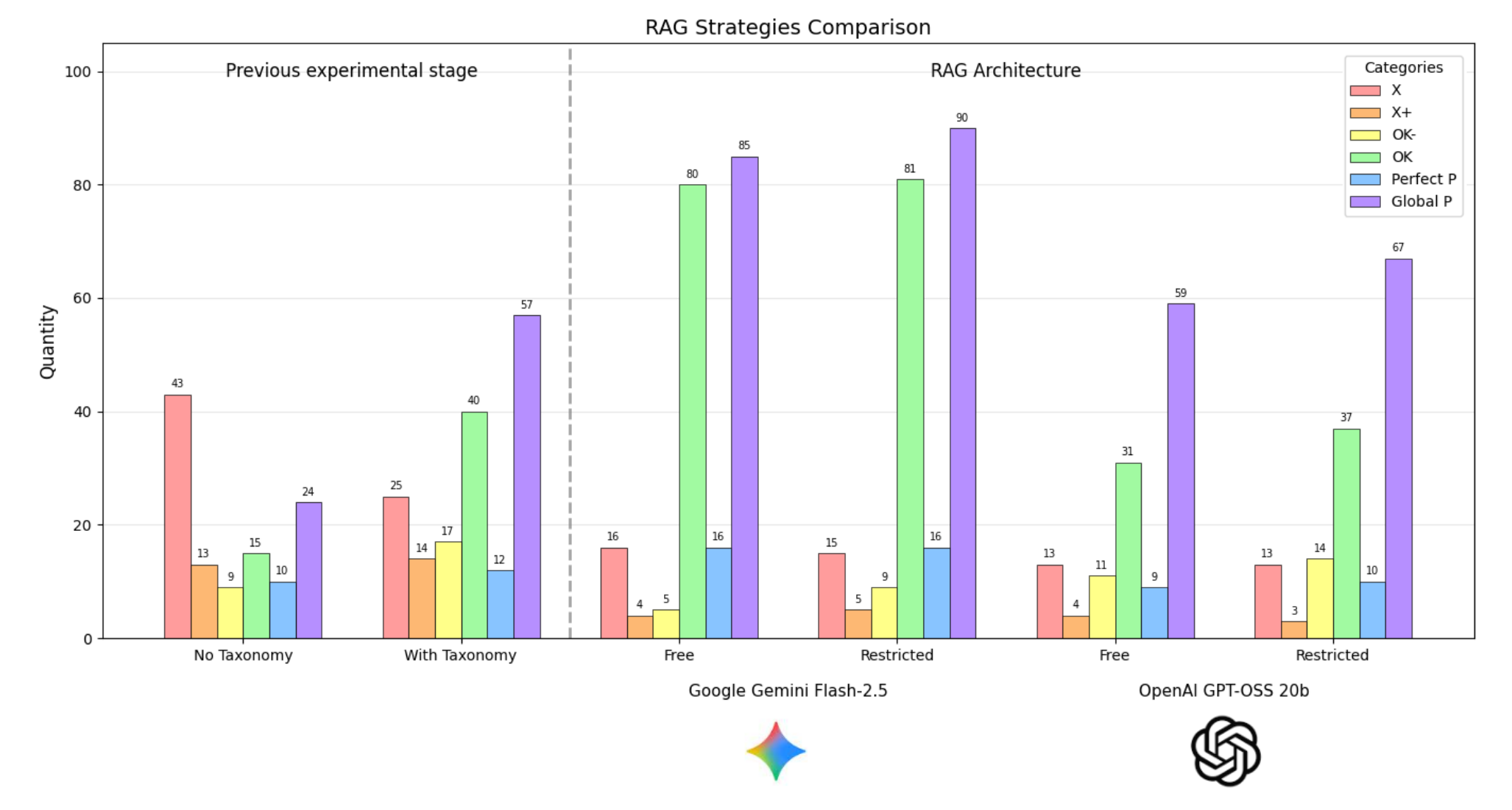}}
    \caption{Evolution of metrics across experimental stages}
    \label{fig:strategy_comparison}
\end{figure}

Regarding experimental correctness and the proportion of invocation failures, we can note that Google Gemini Flash-2.5 obtained an average of 4 scenarios with empty, invalid, or no refactored code responses per test run (15\%), while OpenAI Gpt-oss-20b averaged 16 (61\%). This suggests that OpenAI Gpt-oss-20b proved to be, for our experiment, a more volatile model with greater response variability. Although the free retrieval mode occasionally generates more detailed and descriptive responses, its consistency is low. In contrast, it was observed that OpenAI Gpt-oss-20b frequently presented information gaps, especially in refactoring suggestions, which in several runs remained empty, resulting in incorrect migrated code.

\begin{figure}[htbp]
    \centerline{\includegraphics[width=1\columnwidth]{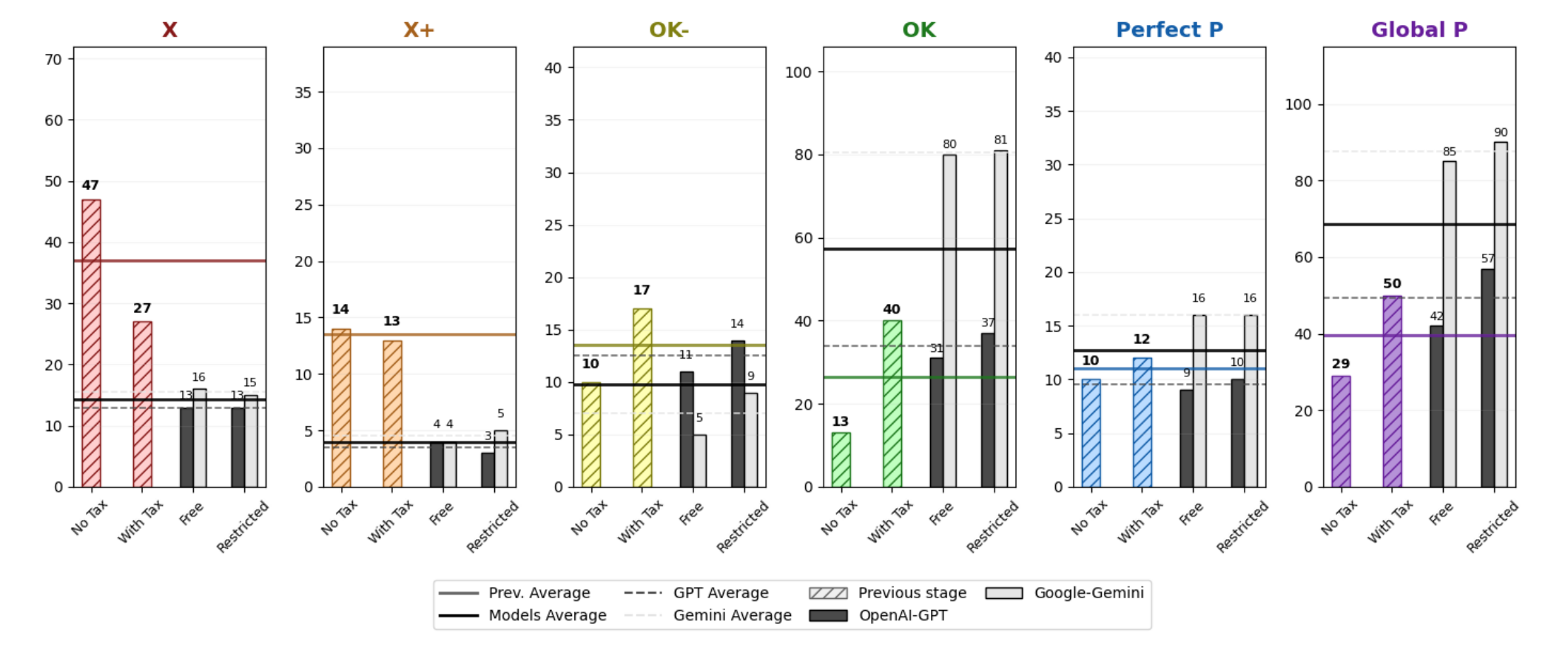}}
    \caption{Evolution of cases according to stoplight metrics}
    \centerline{\tiny{Note: Consider that not all boxes are to the same scale.}}
    \label{fig:spotlight_comparison}
\end{figure}

Along the same lines, if we analyze the progression of the different stoplight metrics with respect to previous stages, comparing them against the average of our current results:
\begin{itemize}
    \item Incorrectly detected scenarios (X, red - see first box from the left in Figure~\ref{fig:spotlight_comparison}): 47 (without taxonomy) and 27 (with taxonomy), reduced to 14.25 on average.

    \item Incorrectly detected scenarios or with inadequate suggestions (X+, orange - see second box from the left in Figure~\ref{fig:spotlight_comparison}): 14 (without taxonomy) and 13 (with taxonomy), reduced to 4 on average.

    \item Scenarios with suggestions requiring minor improvements (OK-, yellow - see third box from the left in Figure~\ref{fig:spotlight_comparison}): 10 (without taxonomy) and 17 (with taxonomy), with an overall average of 9.75.

    \item Correctly detected scenarios with appropriate suggestions (OK, green - see fourth box from the left in Figure~\ref{fig:spotlight_comparison}.): 13 (without taxonomy) and 40 (with taxonomy), increased to 57.25 on average.
\end{itemize}

In summary, the results demonstrate a substantial improvement in model precision through the proposed architecture. Notably, critical errors (X scenarios) were drastically reduced from 47 to an average of 14.25, representing a remarkable decrease in incorrect detections. Concurrently, the system's ability to generate fully satisfactory responses (OK scenarios) saw a remarkable increase, rising from 13 to an average of 57.25 successful cases, where the Gemini model excels in identifying complex refactoring patterns.

In conclusion, there is evidence of improvement that not only validates the use of our taxonomy but also our architecture, especially considering the increase in appropriately detected refactoring scenarios. This stepwise improvement was more pronounced for the Google Gemini Flash-2.5 model, although the trend holds for both. See Figure~\ref{fig:strategy_comparison}.

\section{Conclusion and Future Work}

This work contributes an automated taxonomy integrated into a RAG architecture to optimize semantic database retrieval, alongside an experimental validation demonstrating enhanced detection and accuracy in refactoring suggestions with a direct impact on code quality. Additionally, it introduces an automated, extensible, and modular pipeline that ensures the interoperability and replicability of the experimental environment. The results confirm that AI-based tools can effectively raise the abstraction level in QSE. By providing LLMs with automatically generated, structured domain knowledge integrated into a semantic base, accuracy and quality in mitigating API obsolescence are significantly improved. Consequently, development teams can overcome technical barriers and adopt best practices while decoupling themselves from specific QDK architectural decisions. This approach fosters technological independence and promotes interdisciplinarity within the field.

Among the lines of research and possible future developments, several directions are contemplated:
\begin{itemize}
    \item \textbf{Expansion of the knowledge base}, this involves extending the knowledge base by incorporating additional sources — such as Qiskit changelogs, official migration guides.
    
    \item \textbf{Experimental representativeness and code granularity}, the incorporation of automated pipelines would enable migration from a test base of synthetic Python codes, generated in a controlled manner, toward the integration of projects extracted from public repositories. From the perspective of an integrated tool, managing realistic projects instead of isolated snippets is unavoidable.
    
    \item \textbf{Automated evaluation}, advancing the development of pre- and post-execution stages by incorporating syntactic and semantic validations, automated tests, graph generation, and reporting, for experimental scalability and reliability. 
    
    \item \textbf{Quality metrics}, defining quality metrics that allow for precise comparisons regarding code migration.
    
    \item \textbf{Model granularity}, this involves evaluating the performance of models specifically trained for one ecosystem or specialized in code generation tasks.

\end{itemize}

Considering the interrelation with quantum software engineering (QSE), both the taxonomy and the proposed architecture can be extended to incorporate specific, targeted information sources. The goal is to build a tool that covers functionalities beyond code migration.


\printbibliography

\end{document}